# Assessing Machine Learning Approaches to Address IoT Sensor Drift

Haining Zheng
ExxonMobil Research and Engineering Company
Annandale, NJ 08801
haining.zheng@exxonmobil.com

Antonio R. Paiva
ExxonMobil Research and Engineering Company
Annandale, NJ 08801
antonio.paiva@exxonmobil.com

## ABSTRACT

The proliferation of IoT sensors and their deployment in various industries and applications has brought about numerous analysis opportunities in this Big Data era. However, drift of those sensor measurements poses major challenges to automate data analysis and the ability to effectively train and deploy models on a continuous basis. In this paper we study and test several approaches from the literature with regard to their ability to cope with and adapt to sensor drift under realistic conditions. Most of these approaches are recent and thus are representative of the current state-of-the-art. The testing was performed on a publicly available gas sensor dataset exhibiting drift over time. The results show substantial drops in sensing performance due to sensor drift in spite of the approaches. We then discuss several issues identified with current approaches and outline directions for future research to tackle them.

## 1 Introduction

One of the main motivating factors for the broad deployment of Internet of Things (IoT) devices has been the desire to more broadly instrument processes and activities [1-3]. The deployment of IoT sensors allowing for measurements with higher spatial resolution. In turn, the information provided by these sensors allows for more accurate collection of monitoring information, faster response to events, more effective use of automated analytics, and development of machine learning (ML) [4-6] models. Their deployment has enabled significant progress recently and proven successful across different industries, including a number of applications in the energy industry from upstream production prediction, midstream transportation optimization, to downstream product manufacturing [7-11]. Hence, the usefulness of IoT sensors hinges on the quality of sensor information that they collect and its ability to enable subsequent processing and analysis.

A crucial challenge in many sensor technologies is the issue of sensor drift. The issue can arise because of multiple aspects of the natural interaction with the environment. Over time, sensors can be disturbed by all sorts of environmental conditions and contaminants (e.g., by dust) or experience "aging" due to physical, thermal, or electrical stresses [12-14]. While sensors deemed critical are typically cleaned and recalibrated frequently, the high number of IoT sensors will often mean that they have to endure longer time periods without cleaning or recalibration, which exacerbates the effects of sensor drift. Moreover, note that while developments in sensor technologies may help, they cannot fix all of the reasons for sensor drift. Hence, the application of approaches to cope with and automatically correct for drift are of paramount importance to ensure the accuracy of measurements and the effectiveness of subsequent processes using them.

A number of approaches have been proposed in the literature to try to cope with and adapt to sensor drift. This includes several statistical methods to learn transformations such that sensor drift on the distribution of measurements/features is removed or at least mitigated [14-19]. These approaches are appealing due to their low computational complexity but make significant assumptions about the mode by which drift occurs and thereby changes the data distribution. Other approaches attempt to explicitly retrain the ML models to correct for drift. This can be done in a semi-supervised manner, using self-training for example [21-23], or in a supervised manner [24]. In particular, and in the context of IoT data streams, the Optimized Adaptive Sliding Windowing (OASW) approach has been recently proposed by Yang and Shami [24] to actively detect drift and, in response, retrain an ML model. Unfortunately, the adaptation or retraining requires the ground-truth labels/values become available immediately, or very shortly thereafter. However, this information is not available in practice for sensor measurements, and those from IoT sensors in particular.

## 2 Formulations for addressing sensor drift

In this section, we start by clearly defining the problem and stating what we believe are necessary considerations for realistically addressing sensor drift in IoT devices deployed at large. We then discuss several closely related machine learning paradigms that are particularly relevant to addressing this issue.

## 2.1 Problem statement

Sensor drift refers to the change in response or raw measurements over time. The changes may be the result of multiple factors. These include sensor aging because of physical, thermal, or electrical stresses, contamination due to chemical reactions or other disturbances, or other environmental variations due to temperature or humidity, for instance. These factors result in variations of the sensor's response over time, meaning that the data or derived features will exhibit a different distribution. This change will often be gradual but abrupt changes are also possible.

As previously mentioned, IoT sensor measurements are often acquired to enable learning and applying machine learning models. However, these models are learned with respect to the distribution of the features in the training data. Hence, when the feature space distribution of the data changes, there will be a mismatch between the model's representation and the actual data, resulting in decreased performance. The goal here is to explore methods that can either help adapt the learned models or correct for the changes to the data distribution.

The challenge with IoT sensors is that the adaptation must occur without access to ground-truth information after the sensor has drifted. The ground-truth information, in the form of classification labels in the context of this study, are assumed available only during an initial calibration period. If no correction or adaptation is performed then, as mentioned, the sensing performance will likely deteriorate over time. Hence, the goal of this study was to ascertain how much the drop in performance could be alleviated by using sensor drift approaches.

## 2.2 Related machine learning paradigms

The problem of learning models to adapt to changes of the data distribution has been considered under different paradigms in the machine learning community. These include concept drift, domain adaptation, transfer learning, and semi-supervised learning.

Like the problem we described, concept drift [25] pertains to the development of approaches to detect and adapt to changes in the data distribution. The methods are developed primarily to tackle changes in data streams, as one tends to encounter from IoT devices [24]. However, the problem settings considered under concept drift assume that supervised information becomes available immediately, or very shortly thereafter, making a prediction. If that is the case, then one can retrain the model once statistically significant drops in performance are detected or seek to continuously adapt the model [26]. For IoT sensors, however, ground-truth information is not available after the initial calibration and thus concept drift approaches cannot be practically used.

Domain adaptation and transfer learning also aim to tackle the problem of learning ML models in the presence of changing data distributions [27]. The aim is to leverage the data in a source domain, for which ground-truth information is available, to learn the ML model, while utilizing the target domain to adapt the model such that it might perform robustly in spite of the changes in the feature space distribution. With regard to our problem statement, this means that the data during calibration acts as the source domain and the (unlabeled) data after sensor drift as the target domain.

Clearly, this is a paradigm that is well suited to the sensor drift problem and several of the approaches studied herein adopt that adaptation strategy [18-20].

Yet another way to approach the issue of sensor drift is in the context of semi-supervised learning (SSL) [28]. Like domain adaptation, in SSL the goal is to utilize both labeled and unlabeled data to characterize the feature space distribution, and the ground-truth labels to constrain the inferences to the unlabeled data. Self-training will be considered herein as an SSL approach to tackles sensor drift. It is worth noting that, unlike the other paradigms in this section, SSL does not consider the temporal nature of sensor drift and aims to correctly classify all of the data. Hence, it could be argued that it is attempting to address a harder problem. Domain adaptation in contrast is focused on inference on the target domain (i.e., data with sensor drift).

## 2.3 Other considerations

A large number of approaches have developed under the paradigms overviewed in the previous section. IoT devices have typically very limited computational capabilities and even power restrictions, which constrains the approaches that one can consider. Hence, while there are many domain adaptation or SSL methods which promise better accuracy or robustness (e.g., [29-30]), they are often kernel or graph based and thus have computational requirements beyond the capabilities of most IoT devices. For these reasons, in this study we restricted ourselves to approaches with computational requirements conducive to broad deployment in IoT sensor networks.

## 3 Approaches

### 3.1 Self-training

Self-training is a semi-supervised learning approach. Accordingly, it uses the labeled data during calibration to learn a classification model, while leveraging the feature space distribution for the unlabeled data to adapt. Self-training does this by iteratively generating pseudo-labels for the unlabeled data, from which target data points classified with high confidence are added to the training dataset and used to retrain the model. The level of confidence above which points are added to the training data is a parameter. This approach has been shown to adapt to gradual changes in the data distribution [21, 22].

Self-training has recently been demonstrated to be a highly effective technique for transfer learning and domain adaptation in a number of applications [21-23]. However, as has been previously reported [21], self-training can be prone to amplifying and reinforcing the model errors on the unlabeled data, especially as time progresses and those errors accumulate.

### 3.2 Domain regularized component analysis (DRCA)

Domain regularized component analysis (DRCA) [18] is a domain adaptation technique specifically proposed to tackle the sensor drift problem. The DRCA method is motivated by transfer learning, and the difference of probability distribution between source data and target data incurs the failure of machine learning in data mining. It searches a linear transformation that maps both the "source" and "target" domains to a common subspace. This subspace is assumed to be orthogonal to the difference between the means in each domain or, to put differently, that the change in means is a major drift direction. Then the labels, available for the source domain data, are used to learn a classifier in the transformed space. The DRCA subspace projection is a generalized PCA synthesis and easily solved by Eigen-decomposition. It has two main parameters: the choice of the number of projection components and a trade-off parameter between the covariance matrices of the source and target domains. Previously published DRCA results on synthetic data and real datasets suggested it to be an efficient and effective anti-drift method [18].

### 3.3 Local discriminant subspace projection (LDSP)

A related method, is Local Discriminant Subspace Projection (LDSP) [19], also known as Discriminative Domain Regularized Component Analysis (D-DRCA) [20]. LDSP augments DRCA by further considering the discriminative ability of the transformation projections with respect to the source domain data. The augmented objective uses the source data label information and simultaneously minimizes the within-class variance of the projected source samples and maximizes the between-class difference, as utilized in the Fisher linear discriminant criterion [4]. In other words, the label information is exploited to avoid overlapping of samples with different labels in the subspace. In addition to the parameters of DRCA, LDSP introduces two additional parameters weighing the within-class and between-class matrices into the training objective. Past experiments on two sensor drift datasets have shown the effectiveness of the approach.

**Table 1: Dataset batch numbers and corresponding month and number of data points.**

| Batch ID | Month ID | Number of Data Point |
|---|---|---|
| 1 | 1, 2 | 445 |
| 2 | 3, 4, 8, 9, 10 | 1244 |
| 3 | 11, 12, 13 | 1586 |
| 4 | 14, 15 | 161 |
| 5 | 16 | 197 |
| 6 | 17, 18, 19, 20 | 2300 |
| 7 | 21 | 3613 |
| 8 | 22, 23 | 294 |
| 9 | 24, 30 | 470 |
| 10 | 36 | 3600 |

## 4 Gas sensor array drift dataset

These approaches are demonstrated using a Gas sensor array drift dataset [12]. This dataset contains 13910 measurements from 16 chemical sensors utilized in simulations for drift compensation in a discrimination task of 6 gases at various levels of concentrations. The dataset covers a duration of 36 months in a gas delivery platform facility situated at the ChemoSignals Laboratory in the BioCircuits Institute, University of California San Diego The dataset comprises recordings from six distinct pure gaseous substances, namely Ammonia, Acetaldehyde, Acetone, Ethylene, Ethanol, and Toluene, each dosed at a wide variety of concentration values ranging from 5 to 1000 ppmv. A summary of the dataset can be found in Table 1.

Each measurement produced a 16-channel time series, from which each time series was converted into 2 steady state features and 6 dynamic features. This resulted into 128 features for each data point. The steady-state feature is defined as the difference of the maximal resistance change and the baseline when the chemical vapor is present in the test chamber and its normalized version was calculated by the ratio of the maximal resistance and the baseline. The sensor dynamic features presented the increasing and decaying transient portion of the sensor response during the entire measurement procedure under controlled conditions. Three different smoothing factors were employed for the exponential moving average of the increasing and decaying processes.

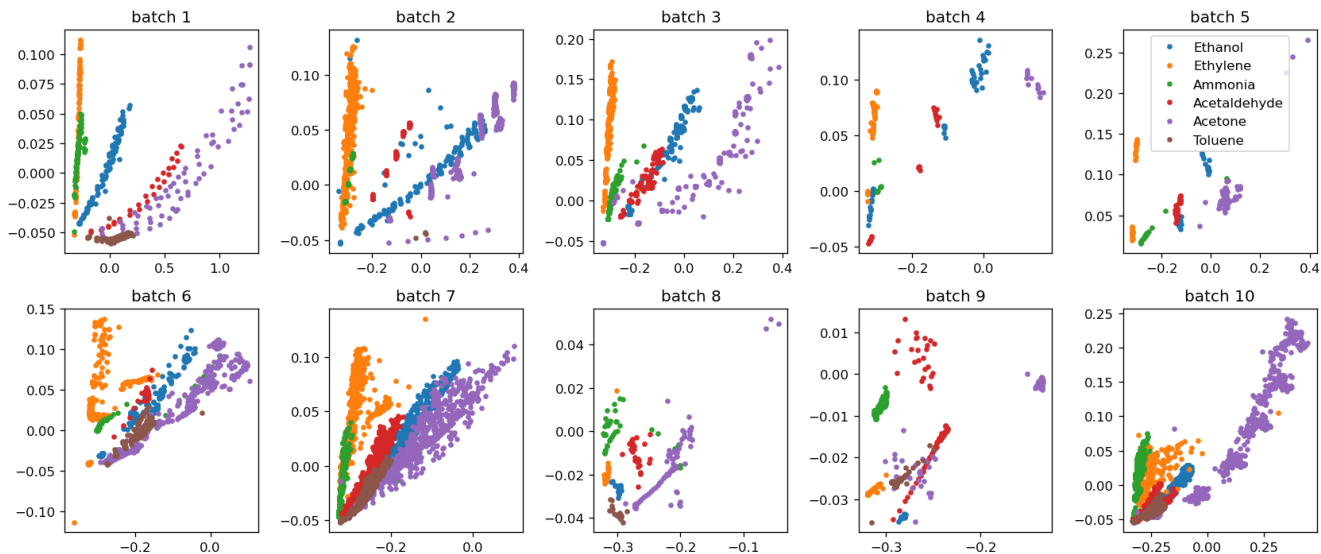

**Figure 1: Unsupervised visualization of feature space distribution for each batch using the first two PCA principal components.**

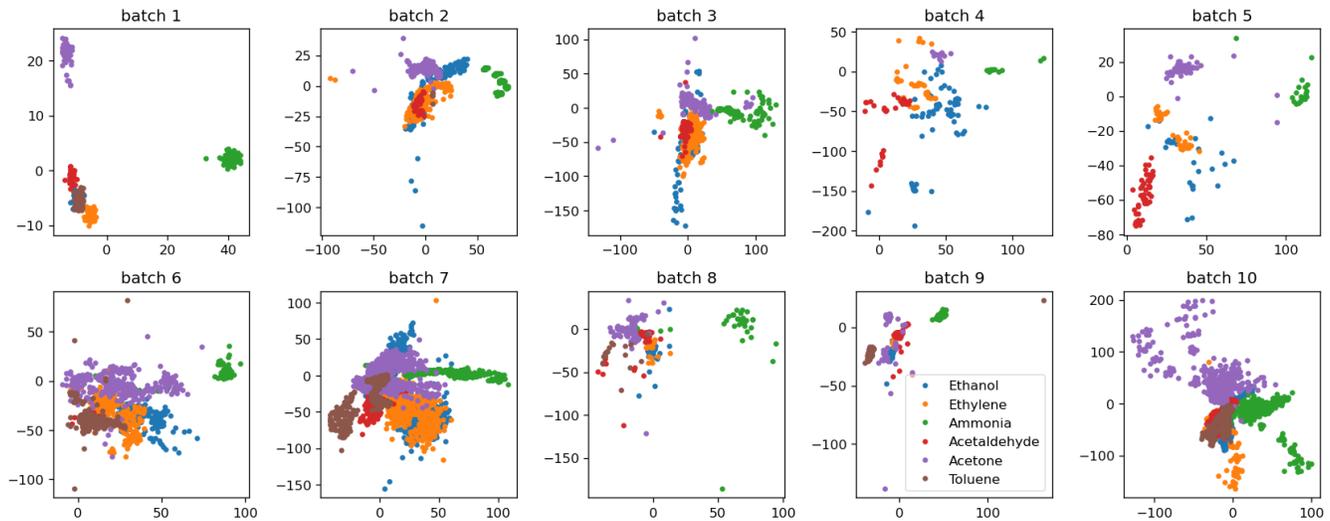

**Figure 2: Projections of each batch onto the first two components of Linear Discriminant Analysis computed using only the label data from batch 1.**

A visualization of feature space distribution for each batch using the first two PCA principal components, computed from batch 1, is given in Figure 1. Similarly, Figure 2 shows the linear discriminant projection analysis (LDA) using the labels from batch 1. Sensor drift can be clearly observed between different batches from the projections. Note, for example, the change in relative position in the PCA projections between Ethanol and Acetaldehyde data points, or between Ethylene and Ammonia data points. Or the significant increase in overlap of the classes in the LDA projections.

In the following, the dataset will be considered in the context of a classification problem. That is, we aim to identify the gas from the measurements. For the subsequent analysis, we further used the smallest and largest values of each feature in batch 1 to shift and scale each feature across all batches. For batch 1, this normalized each feature separately to the [0, 1] interval. In spite of this normalization, it must be noted that the data still had tremendous dynamic range variations, with features values in batch 2 as low as -57.8 and as high as 2624 in the extreme. Large dynamic range values are known to generally create challenges for methods based on the data's second-order statistics, such as DRCA and LDSP, but represent the possibility of these methods encountering outlier or anomalous values. Nonetheless, it was observed that performing this "normalization" helped training and yielded much better results of the different classifiers.

## 5 Results

As discussed in sections 2 and 3, we considered Domain regularized component analysis (DRCA), local discriminant subspace projection (LDSP) and self-training as promising methods to tackle sensor drift and that are widely deployable in IoT devices.

These approaches were tested on Gas sensor array drift dataset described in section 4. The problem was formulated as a multi-class classification task for the six distinct pure gaseous substances: Ammonia, Acetaldehyde, Acetone, Ethylene, Ethanol, and Toluene. All experiments shown here used logistic regression (LR) with $l_2$ regularization as the classifier. We also tried other classifiers, including Linear Discriminant Analysis, LightGBM, and SVMs with linear, polynomial, or RBF kernels, but they yielded lower classification performance. It must be emphasized that the crucial issue here is robustness to sensor drift, and it would appear that the simplicity and strong regularization in LR were beneficial.

In the experiments batch 1 is considered as the source domain source domain and all subsequent batches as the target domains. In other words, per the problem statement in section 2.1, batch 1 is taken to be the calibration data and only those labels are available to train the classifiers. The labels from subsequent batches are used only to evaluate the performance of the different approaches.

The parameters of DRCA and LDSP were chosen to obtain the best average classification accuracy. The number of transformation components was kept as 127, since this maximally preserved the information for the classifiers and yielded the best results. Note that the removal of the direction between the means of the batches reduces the dimensionality of the space by one. The trade-off parameter between covariance matrices was set as $\lambda = 0.1$ in both cases, and the weight of the within-class and between-class matrix

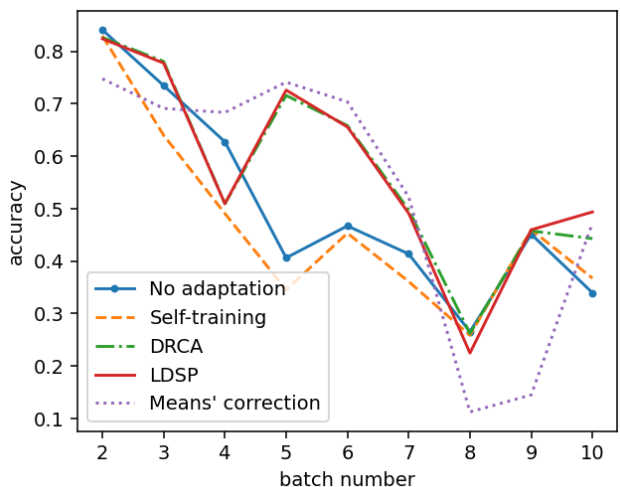

**Figure 3: Comparison of classification accuracy of different approaches after initial calibration in Batch 1.**

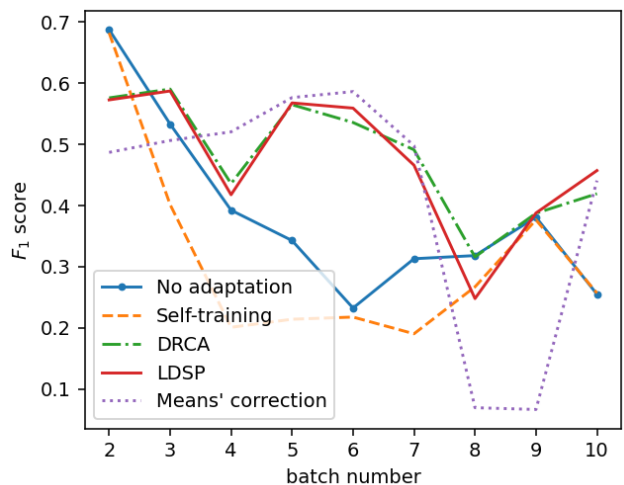

**Figure 4: Comparison of $F_1$ score results of different approaches, as in Figure 3.**

terms in the LDSP objective as $\kappa = 10$ and $\mu = 100$, respectively (see eq. 8 in [19]).

The classification accuracy and $F_1$ score (multiclass averaged) of each of the approaches are shown in Figures 3 and 4, respectively. As a baseline, the "No adaptation" case corresponds to the situation in which the classifier was trained on batch 1 and applied without change or any form of adaptation to the subsequent batches. As expected, sensor drift in the measurement causes significant drops in the classification performance over time if no adaptation is performed. To our surprise, self-training performed even worse than no adaptation. Even when restricting to data points with high classification confidence ($\geq 0.99$), as shown, it would seem that self-training just accentuated and reinforced the model errors in subsequent batches.

DRCA and LDSP performed much better than no correction, especially for batches 5 through 7. This indicates that indeed using the feature distribution of the unlabeled data points is beneficial to adapt and cope in the presence of sensor drift. It is noteworthy that, for both of these methods, there was substantial variation on the classification accuracy of each batch for different values of the parameters. Here, the parameters were constant for all batches and chosen to achieve the best average classification accuracy, which in itself depicts an optimistic test situation. Still, while it is possible to obtain better results for each batch by optimizing the approach's parameters, doing so would require the labels of each batch, which of course is not at all practical.

Finally, since both the DRCA and LDSP objectives restricted their transformation to subspaces that removed the direction between the means of each domain (i.e., batch), a "Means' correction" case is considered. As its name implies, we simply centered the data from each batch. In this case, this very simple procedure maintained a robust performance of 70-75% until batch 6. That's a period of 18 months. It actually has the best accuracy on batches 4-7, but it performs much worse afterward. These results are shown here not to suggest using this approach but to highlight the significant challenges in addressing sensor drift and the significant opportunities to more effectively address this issue.

Overall, there seem to be two main issues with above the DRCA and LDSP approaches considered on our study: (1) they are highly sensitive to choice or the regularization/weighting parameters, and their optimal value makes strong assumptions on the relation between the subspace of transformations and the direction(s) of drift, and (2) they are sensitive to outlier directions. To contemplate the second issue, consider that in Figure 2 we restricted the plotting to points with absolute projection values under 200. Otherwise, there would be a handful of points with very large values, most notably in batch 2. Although the LDA projection using the labels from batch does not have this problem (see Figure 5), drift adaptation approaches need to optimize against such directions without access to the labels. Approaches which optimize with regard to second-order statistics are likely to be challenged. These observations were not affected by the feature normalization.

## 6 Conclusions and future work

This paper considered three recent approaches from the literature and tested them with regard to their ability to adapt and cope to sensor drift under realistic conditions. The result of our tests clearly show significant gaps in performance, and opportunities for improvement, in the methods to address IoT sensor drift under realistic conditions.

As discussed at the end of the previous section, we believe that future research needs to consider approaches with more robust statistics or that more comprehensively utilizes the feature distribution in subsequent batches. A major consideration however is how to balance those needs with the limited computational capabilities in IoT devices. The other aspect would be for future approaches to have methods to tune their parameters under realistic scenarios, or demonstrate robust performance across a broad range of their values.

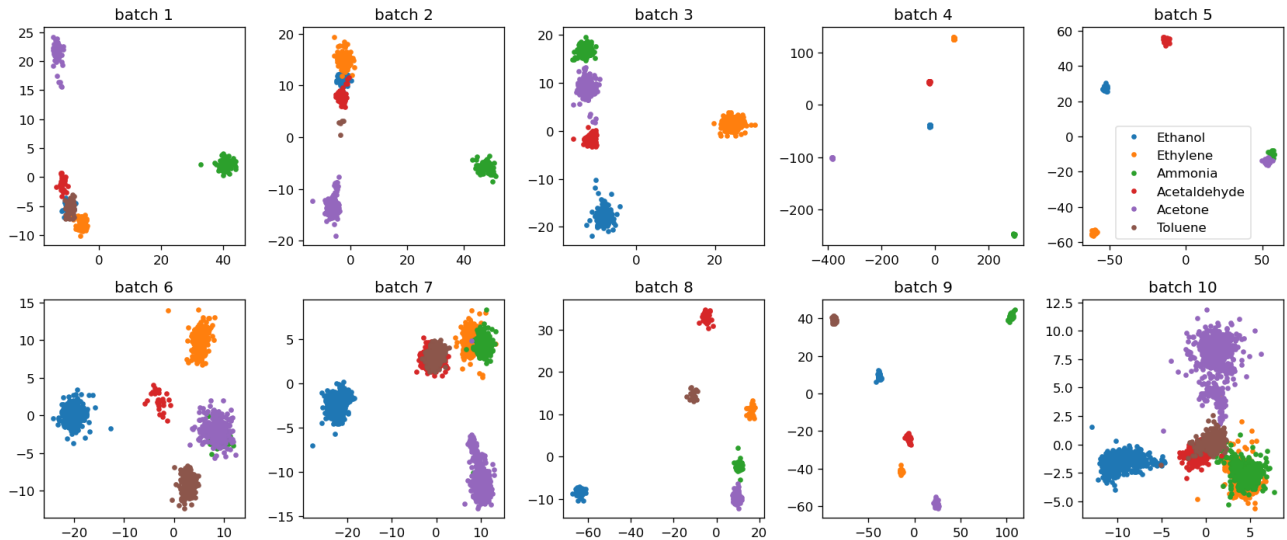

**Figure 5: Projections onto the first two components of Linear Discriminant Analysis applied to each batch.**

## Acknowledgements

The authors would like acknowledge the support from Sasha Mitarai, Prasenjeet Ghosh and Myun-Seok Cheon.